\documentclass[12pt]{article}
\usepackage{amsfonts,amsmath,amssymb}
\usepackage{graphicx}

\newfont{\twelvemsb}{msbm10 scaled\magstep1}
\newfont{\eightmsb}{msbm8}
\newfam\msbfam
\textfont\msbfam=\twelvemsb \scriptfont\msbfam=\eightmsb
\catcode`\@=11
\def\Bbb{\ifmmode\let\next\Bbb@\else
\def\next{\errmessage{Use \string\Bbb\space only in math mode}}\fi\next}
\def\Bbb@#1{{\fam\msbfam{{#1}}}}

\newcommand{\be}{\begin{equation}}
\newcommand{\ee}{\end{equation}}
\newcommand{\ba}{\begin{eqnarray}}
\newcommand{\ea}{\end{eqnarray}}

\newcommand{\m}{\mathcal}
\newcommand{\ti}{\tilde}

\begin{document}

\sloppy
\renewcommand{\thefootnote}{\fnsymbol{footnote}}
\newpage
\setcounter{page}{1} \vspace{0.7cm}
\vspace*{1cm}
\begin{center}
{\bf On the scattering over the GKP vacuum}\\
\vspace{.6cm} {Davide Fioravanti $^a$, Simone Piscaglia $^{a,b}$} \\
\vspace{.3cm} $^a$ {\em Sezione INFN di Bologna, Dipartimento di Fisica e Astronomia,
Universit\`a di Bologna} \\
{\em Via Irnerio 46, Bologna, Italy}\\
\vspace{.2cm}  $^b$ {\em Centro de F\'isica do Porto and Departamento de F\'isica e Astronomia,
Universidade do Porto} \\
{\em Rua do Campo Alegre 687, Porto, Portugal}\\
\vspace{.5cm} {Marco Rossi $^c$}
\footnote{E-mail: fioravanti@bo.infn.it, piscagli@bo.infn.it, rossi@cs.infn.it} \\
\vspace{.3cm} $^c${\em Dipartimento di Fisica dell'Universit\`a
della Calabria and INFN, Gruppo collegato di Cosenza} \\
{\em Arcavacata di Rende, Cosenza, Italy} \\
\end{center}
\renewcommand{\thefootnote}{\arabic{footnote}}
\setcounter{footnote}{0}
\begin{abstract}
{\noindent By converting the Asymptotic Bethe Ansatz (ABA) of ${\cal N}=4$ SYM into non-linear integral equations, we find 2D scattering amplitudes of excitations on top of the GKP vacuum. We prove that this is a suitable and powerful set-up for the understanding and computation of the whole S-matrix. We show that all the amplitudes depend on the fundamental scalar-scalar one.}
\end{abstract}
\vspace{6cm}

\newpage

\section{Introduction}
\setcounter{equation}{0}

In integrable system and condensed matter theories the study of the scattering of excitations over the antiferromagnetic vacuum is at least as much important as that over the ferromagnetic one ({\it cf.} one of the pioneering papers \cite{Faddeev:1981ft} and its references). Often the excitations over the ferromagnetic vacuum are called magnons, as well as those over the antiferromagnetic one kinks or solitons or spinons. Also, two dimensional (lattice) field theories (like, for instance, Sine-Gordon) are often examples with an antiferromagnetic vacuum \cite{Faddeev:1979gh}. If we wish to parallel this reasoning in the framework of the Beisert-Staudacher  asymptotic ({\it i.e.} large $s$) Bethe Ansatz (ABA) for ${\cal N}=4$ SYM \cite{Beisert-Staudacher}, we are tempted to choose, as antiferromagnetic vacuum, the GKP long ({\it i.e.} fast spinning) $AdS_5$ string solution \cite {GKP}. According to the AdS/CFT correspondence \cite {MGKP2W}, the quantum GKP string state is dual to a single trace twist two operator of ${\cal N}=4$ (at high spin); thus let us consider two complex scalars, $Z$, at the two ends of a long series of $s$ (light-cone) covariant derivatives, $D_+$. Then, excitations of the GKP string correspond to insertions of other operators over this Fermi sea, thus generating higher twist operators. More precisely, operators associated to one-particle $\varphi$ states are built as
\be
\mathcal{O'}=Tr\, Z D_+^{s-s'} \varphi D_+^{s'} Z + \dots \quad .
\ee
The set of lower twist (twist three) excitations includes $\varphi=Z$, one of the three complex scalars;  $\varphi=F_{+\bot}, \bar{F}_{+\bot}$, the two components of the gluon field; $\varphi= \Psi_+$, $\bar{\Psi}_+$, the $4+4$ fermions, respectively. All these fields are the highest weight of a precise representation of the residual $so(6)\simeq su(4)$ symmetry of the GKP vacuum: the scalar of the vector ${\bf 6}$, the fermions of the ${\bf 4}$ and ${{\bf \bar 4}}$, respectively, and the gluons of the ${\bf 1}$ representation. All these features and the exact dispersion relations for these excitations in the different regimes have been studied recently in deep and interesting detail by Basso \cite{Basso}. Now, as for the scattering, we shall consider at least two particles states (twist-$4$), namely
\be
\mathcal{O''}=Tr\, Z D_+^{s-s_1-s_2} \varphi_1 D_+^{s_1} \varphi_2 D_+^{s_2} Z+ \dots \quad .
\ee
This situation was already analysed in partial generality in the case when both the excitations are identical $\varphi_1=\varphi_2=Z$ \cite{FRO6}.
Importantly, an impressive recent paper \cite {BSV}  proposed a non-perturbative approach to 4D scattering amplitudes in ${\cal N}=4$ SYM by using as building blocks the 2D scattering amplitudes we wish to compute here: we will see some non-trivial checks of their conjectures.

Computing the scattering matrix has a long history (see, for instance, \cite{Doikou:1999xz} and references therein). From this wide literature we can argue that an efficient method of computation rests on the non-linear integral equation (for excited states) \cite {FMQR}. In fact, the same idea of counting function gives a quantisation condition which can be interpreted as (asymptotic) Bethe Ansatz, defining the scattering matrix (elements). In this note we will use this strategy to provide general formul{\ae} for the scattering amplitudes between the aforementioned excitations. This should give the non-trivial part (in front) of the scattering matrices, the so-called scalar factor, being the matrix structure fixed by the aforementioned residual symmetry representations. Remarkably, all the scattering phases (eigenvalues) are expressible in terms of the scalar-scalar one. Moreover, we evaluate one loop and strong coupling limits of (scalar-scalar and) gluon-gluon scattering amplitudes and find confirmation of the conjecture of \cite{BSV}.

\section{Scalar excitations}
\setcounter{equation}{0}
A $sl(2)$ state of  $L$ ($=$twist) scalars and $s$ ($=$spin) derivatives is described by the counting function
\be
Z(u)=L \Phi (u)-\sum _{k=1}^{s}
\phi (u,u_k) \, , \label {Z}
\ee
where $\Phi (u)=\Phi _0(u)+\Phi _H (u) \, , \quad \phi (u,v)=\phi _0(u-v)+\phi _H (u,v)$ , with
\ba
\Phi _0(u) &=&-2 \arctan 2u \, , \quad \Phi _H(u)=-i \ln \left ( \frac {1+\frac {g^2}{2{x^-(u)}^2}}{1+\frac {g^2}{2{x^+(u)}^2}} \right )\, ,  \label {Phi} \\
\phi _0(u-v)&=& 2\arctan (u-v) \, ,\quad \phi _H(u,v)=-2i \left [ \ln \left ( \frac {1-\frac {g^2}{2x^+(u)x^-(v)} }{1-\frac {g^2}{2x^-(u)x^+(v)}} \right )+i\theta (u,v)\right] \, , \label {phi}
\ea
$\theta (u,v)$ being the dressing phase \cite {BES} and $ x(u)=\frac{u}{2}\left[1+\sqrt{1-\frac{2g^2}{u^2}}\right]$, $ x^\pm(u)=x(u\pm\frac{i}{2})$.

\noindent
{\bf One loop or learning the art.--} Let us start by reviewing the one loop formulation, in order to outline the main steps of our procedure and to
enucleate several conceptual features that will be encountered again at all-loops.
In the one loop case, the counting function for the twist sector is $Z_0(u)= \Phi _0(u) - \displaystyle{\sum _{k=1}^{s}}
\phi _0(u,u_k)$.
We remark that by its definition $Z_0(u)$ is a monotonously decreasing function.
$L$ holes, in positions $x_h$, $h=1,...,L$ are present. Two of them, in positions $x_1$, $x_L$ are external to Bethe roots (i.e. $x_1<u_k<x_L$); internal or 'small' holes occupy positions we denote with $x_2,...x_{L-1}$.
At one-loop, the counting function satisfies the non-linear integral equation \cite {FRS}
\be
Z_0(u)=\frac{L}{i} \ln \frac{\Gamma \left (\frac{1}{2}+iu\right )}{\Gamma \left (\frac{1}{2}-iu\right )}+i\sum _{h=1}^L \ln \frac{\Gamma (1+iu-ix_h)}{\Gamma (1-iu+ix_h)}+ \int _{-\infty}^{+\infty} \frac{dv}{\pi} [ \psi (1+iu-iv)+ \psi (1-iu+iv) ]L_0(v) \, , \quad \label {1loopZ}
\ee
where $L_0(u)= {\mbox {Im}} \ln [1+(-1)^{L} \, e^{iZ_0(u-i0^+)}]$. We want to study excitations on top of the GKP string. The GKP string is dual to the large spin limit of the twist two operator with only the two external holes, corresponding to the two $Z$ scalars. Then, excitations arise when $L>2$ and are described by the $L-2$ internal holes, corresponding to insertions of $L-2$ fields $Z$. In the high spin limit and within accuracy $O((\ln s)^0)$, equation (\ref {1loopZ}) linearises, because of the expansion $\int _{-\infty}^{+\infty} \frac{dv}{\pi} [ \psi (1+iu-iv)+ \psi (1-iu+iv) ]L_0(v)= -2 u \ln 2  + O \left ( \frac{1}{s^2} \right )$ \cite {FRS}, and we can extract scattering data in the following way. By definition of counting function
\be
(-1)^{1-L}=e^{-iZ_0(x_h)}=e^{iR P_0(x_h)}\prod _{\{h'=2,\ h'\neq h\} }^{L-1} \left (-S_0(x_h,x_{h'})\right) \Rightarrow e^{iR P_0(x_h)}\prod _{\{h'=2,\ h'\neq h\} }^{L-1} S_0(x_h,x_{h'})=1
 \, , \label {Z-pS}
\ee
for $h=2,...,L-1$, where the last equalities represent the Bethe quantisation conditions: $R$ is given the interpretation of the effective length of the closed chain and $P_0(x_h)$ that of the momentum of the $h$-th excitation, so that $R P_0(x_h)$ represents the 'free propagation' phase of this excitation around the chain (the minus sign in front of $iZ_0(x_h)$ is due to our definition of $Z_0(u)$ as a decreasing function). On the contrary, $S_0(x_h,x_{h'})$ allows for the interaction and is given the interpretation of phase change due to the scattering between the $h$-th and the $h'$-th excitation. To ensure unitarity and proper asymptotic behaviour to the scattering, we add and subtract one term in (\ref {1loopZ}):
\ba \label {Z_0}
&&Z_0(u)=
\left [ -i (L-2) \ln \frac{\Gamma \left (\frac{1}{2}+iu \right )}{\Gamma \left (\frac{1}{2}-iu \right )} + i \sum _{h=2}^{L-1} \ln \frac{\Gamma \left (\frac{1}{2}+ ix_h\right )}{\Gamma \left (\frac{1}{2}- ix_h\right )}+
i \sum _{h=2}^{L-1} \ln \frac{\Gamma (1+iu -ix_h)}{\Gamma (1-iu+ix_h)} \right ] + \quad\quad\\
&& \left [ -2i \ln \frac{\Gamma \left (\frac{1}{2}+iu \right )}{\Gamma \left (\frac{1}{2}-iu \right )}-i \sum _{h=2}^{L-1} \ln \frac{\Gamma \left (\frac{1}{2}+ ix_h\right )}{\Gamma \left (\frac{1}{2}- ix_h\right )}
+i \ln \frac{\Gamma (1+iu-ix_L)\Gamma (1+iu-ix_1)}{\Gamma (1-iu+ix_L)\Gamma (1-iu+ix_1)} - 2u\ln 2 \right ] \nonumber \, .
\ea
In the large spin limit we have \cite {BGK}
$x_L=-x_1=\frac{s}{\sqrt{2}}+O\left(s^0\right)$, while  $x_h \sim \frac{1}{\ln s}$,\,\, $2\leq h\leq L-1$\cite {FGR5}. Therefore, the second bracket in (\ref {Z_0}) reduces to
$
-4u\ln s  -2i \ln \frac{\Gamma \left (\frac{1}{2}+iu \right )}{\Gamma \left (\frac{1}{2}-iu \right )} +
O\left(\frac{1}{(\ln s)^2}\right)
$.
When no internal hole is present at all, the first bracket in (\ref {Z_0}) is absent and we identify the effective length of the string and the excitation momentum as $R = 2\ln s $ and $P_0(x_h)=2x_h $, respectively.
On the other hand, to find the scattering matrix involving two scalar excitations, it is convenient to stick to
the $L=4$ case (two internal holes). Now, the extra phase shift, due to the first bracket in (\ref {Z_0})
is interpreted, via (\ref {Z-pS}), as the scattering factor
\be
S_0(x_h,x_{h'})= -\frac{\Gamma \left (\frac{1}{2}-ix_h \right ) \Gamma \left (\frac{1}{2}+ix_{h'} \right ) \Gamma (1+ix_h-ix_{h'})}{\Gamma \left (\frac{1}{2}+ix_h \right )
\Gamma \left (\frac{1}{2}-ix_{h'} \right )\Gamma (1-ix_h+ix_{h'})}    \, , \label {S_0}
\ee
between two internal holes with rapidities $x_h$ and $x_{h'}$. This expression enjoys unitarity and does agree with result (3.8) of Basso-Belitsky \cite{BB}, but seems to be the inverse of (2.13) in \cite{DoreyZhao}.

\medskip

\noindent
{\bf All loops.--} Using results contained in Section 2 of \cite {BFR}, we write for the counting function the following non-linear integral equation
\be
Z(u)=F(u)+2\int _{-\infty}^{+\infty} dv G(u,v) L(v) \, ,
\ee
where the functions $F(u)$ and $G(u,v)$ are obtained after solving the linear integral equations
\be
F(u)=f(u) - \int _{-\infty}^{+\infty} dv \varphi (u,v) F(v) \, , \,\,\,\,\,
G(u,v)=\varphi (u,v)- \int _{-\infty}^{+\infty} dw \varphi (u,w) G(w,v) \, , \label {Geq}
\ee
with
\be
f(u)=L \Phi (u) + \sum _{h=1}^L \phi (u,x_h) \, , \quad \varphi(u,v)=\frac{1}{2\pi} \frac{d}{dv} \phi (u,v) \label {fphidef} \, .
\ee
Equations (\ref {fphidef}) and (\ref {Geq}) entail the sum $F(u)=L \tilde P(u)+\sum \limits _{h=1}^L R(u,x_h)$ of the two functions $R(u,v)$, such that $\frac{1}{2\pi} \frac{d}{dv} R(u,v) = G(u,v)$, and $\tilde P(u)$, solutions respectively of the two linear equations
\be
R(u,v)=\phi (u,v) - \int _{-\infty}^{+\infty} dw \varphi (u,w) R(w,v) \, , \,\,\,\,
\tilde P(u)=\Phi (u) - \int _{-\infty}^{+\infty} dw \varphi (u,w) \tilde P(w) \, . \label {tildePeq}
\ee
Now, we consider the high spin limit, work out the nonlinear term  and - following what we did in the one loop case -
identify the momentum of a hole and the scattering phase between two holes. In fact, the nonlinear term $NL(u)=2\int dv G(u,v) L(v) $ depends on the function $G(u,v)$. By manipulating the second of (\ref {Geq}) in Fourier space and using formula (3.2) of \cite {FRO6}, we arrive at this (approximated) integral equation $\hat NL(k)= - \frac{4\pi \ln 2}{ik} \delta (k)  -
\frac{1}{1-e^{-|k|}} \int _{-\infty}^{+\infty} \frac{dp}{2\pi} \hat \varphi _H (k,p) \hat NL(-p) +O(1/s^2)$, which proves that $NL(u)$ start contributing at order $O(s^0)$. Again, as in the one loop case, we use the quantisation conditions
\be
(-1)^{1-L}=\exp (-i Z(x_h))=e^{iR P(x_h)}\prod _{\{h'=2,\ h'\neq h\} }^{L-1} \left (-S(x_h,x_{h'}) \right )
\ee
to define the momentum of a hole/scalar excitation of rapidity $u$ as the function $P(u)$ such that
\be
-2 \tilde P(u) - R(u,x_L)- R(u,x_1) - \sum _{h=2}^{L-1} \tilde P(x_h) - NL(u) \simeq R \cdot P(u)  \, , \label {mom}
\ee
with $R\simeq 2\ln s$ (since $x_L=-x_1\simeq \frac{s}{\sqrt{2}}$) the effective length of the chain. On the other hand, the scattering factor between two holes of rapidities $u$, $v$ is the function $S(u,v)$ defined by
\be
i \ln \left (-S(u,v)\right )= R(u,v) + \tilde P(u)-\tilde P(v)=\Theta (u,v) \, .
\ee
We remark that the following properties hold
\be
\tilde P(u)=-\tilde P(-u) \, , \quad R(u,v)=-R(-u,-v) \, , \quad R(u,v)=-R(v,u) \label {Rprop} \, .
\ee
In particular, the last property implies unitarity, {\it i.e.} $ \Theta (u,v)=-\Theta (v,u)$.
To compute this phase, we need first its 'reduced' version $\Theta '(u,v)=R(u,v)+\tilde P(u)$, which satisfies, in (double) Fourier space,
\be
\hat \Theta '(k,t)= \hat \phi (k,t)+\hat \Phi(k)2\pi \delta (t) - \int _{-\infty}^{+\infty} \frac{dp}{4\pi ^2} ip \hat \phi (k,p) \hat \Theta '(-p,t) \, ,
\label{theta'}
\ee
upon manipulating and adding equations (\ref {tildePeq}).
In fact, this function enters the even part (in the second variable) of the scattering phase $M(u,v)=\frac{\Theta (u,v)+\Theta (u,-v)}{2} = \frac{\Theta '(u,v)+\Theta ' (u,-v)}{2}$
, whose (double) Fourier transform, by virtue of (\ref{theta'}), satisfies the equation
\ba
\hat M(k,t)&=&\frac{\hat \phi _H(k,t)+\hat \phi _H(k,-t)}{2(1-e^{-|k|})}-2\pi ^2 \delta (t)
\frac{J_0(\sqrt{2}gk)}{ik \sinh \frac{|k|}{2}} + \nonumber \\
&+& \frac{2\pi ^2}{ik} \frac{e^{-|k|}}{1-e^{-|k|}} [\delta (k+t)+\delta (k-t)]
- \int _{-\infty}^{+\infty} \frac{dp}{4\pi ^2}
\frac{ip \hat \phi _H (k,p)}{1-e^{-|k|}} \hat M (-p,t)
   \label {Meq1} \, .
\ea
We observe that $\hat M(k,t)$ enjoys the parity properties
\be
\hat M (k,t)=\hat M (k,-t) \, , \quad \hat M (k,t)=-\hat M (-k,t) \, ; \label {Mpar}
\ee
the first property is true by construction, the second one holds since the function $\hat \phi _H(k,t)+\hat \phi _H(k,-t)$ is an odd function of $k$. Now, the key point is that we can relate $\hat M(k,t)$ to functions we found in the study of high spin twist sector. Let us consider the density corresponding to the first generalised scaling function (i.e. the part of the density proportional to $\ln s \frac{L-2}{\ln s}$, see \cite {FGR1} for details\footnote {In previous literature integral equations are often written by using the 'magic kernel' $\hat K$ \cite {BES}, related to $\hat \phi _H$ by
\be
\hat \phi _H(k,t)+\hat \phi _H(k,-t)=8i\pi ^2 g^2 e ^{-\frac{t+k}{2}} \hat K (\sqrt{2}gk, \sqrt{2}gt) \, , \quad t,k >0 \, . \nonumber
\ee
}):
\be
\hat \sigma ^{(1)}(k)=\frac{\pi }{\sinh \frac{|k|}{2}}[e^{-\frac{|k|}{2}}-J_0(\sqrt{2} gk) ]
+\frac{i k}{1-e^{-|k|}} \int _{-\infty}^{+\infty} \frac{dt}{4\pi ^2}
\hat \phi _H (k,t) \Bigl [ 2\pi + \hat \sigma ^{(1)}(t)  \Bigr ] \, .  \label {sigma(1)}
\ee
Then, consider the density 'all internal holes', which satisfies equation (3.8) of \cite {FRO6}: we formally put $L=3$ and highlight the dependence of the solution of (3.8) of \cite {FRO6} on the position $x$ of the (fictitious) single internal hole
\be
\hat \sigma (k;x)= \frac{2\pi e^{-|k|}}{1-e^{-|k|}} \left (\cos k x  -1 \right ) +  \frac{ik}{1-e^{-|k|}} \int _{-\infty}^{+\infty} \frac{dt}{4\pi ^2}
\hat \phi _H (k,t) \Bigl [ 2\pi (\cos t x -1 )+ \hat \sigma (t;x) \Bigr ] \, . \label {tilSigeqA}
\ee
Fourier transforming with respect to $x$, it is easy to see that
\be
ik \hat M(k,t)=\int _{-\infty}^{+\infty} dx e^{-itx} [ \hat \sigma ^{(1)}(k) + \hat \sigma (k;x)] \Rightarrow
\frac{d}{du} M(u,v)= \sigma ^{(1)}(u)+\sigma (u;v)   \label {M-sigma} \, .
\ee
In order to fix $M$ from (\ref {M-sigma}), we use properties (\ref {Mpar}): we obtain that $M(u,v)=Z^{(1)}(u)+Z(u;v)$,
where $Z^{(1)}(u)$ and $Z(u;v)$ are univocally defined by the conditions
\be
\frac{d}{du}Z^{(1)}(u)=\sigma ^{(1)}(u) \, , \ \ \frac{d}{du} Z(u;v)= \sigma (u;v) \, , \ \ Z^{(1)}(u)=-Z^{(1)}(-u) \, , \ \ Z(u;v) = -Z(-u;v) \, . \label  {Z-cond}
\ee
Now, we easily analyse the odd part of the scattering phase $N (u,v)=\frac{\Theta (u,v)-\Theta (u,-v)}{2}=\frac{R(u,v)- R(u,-v)}{2} - \tilde P(v)$, for which properties (\ref {Rprop}) bring about $M(v,u)=-N(u,v)$. As a consequence, the scattering phase $\Theta (u,v)$ can be expressed in terms of only the function $M$ as
\be
\Theta (u,v)=M(u,v)-M(v,u) \, . \label {ThetaMrel}
\ee
\noindent
{\bf Two strong coupling limits}

Using (\ref  {M-sigma}, \ref {ThetaMrel}), we want to study the strong coupling limit of the holes scattering phase. We analyse two limits: the so-called non-perturbative regime \cite {AM}, in which $g\rightarrow +\infty$, with $u,v\sim 1 $ fixed and the scaling cases, when we first rescale the rapidities, $u=\sqrt{2}g \bar u$, $v=\sqrt{2}g \bar v$, and then send $g\rightarrow +\infty$, with $\bar u$, $\bar v$ fixed. In the latter case, two regimes appear: if $|\bar u|<1$, $|\bar v|<1$, we are in the perturbative regime, if $|\bar u|>1$, $|\bar v|>1$, we are in the giant-hole \cite {DL} regime.

In the non-perturbative regime we use results (3.21,3.22) of \cite {FRO6}, i.e.:
\be
\hat \sigma ^{(1)}(k) + \hat \sigma (k;x) \rightarrow \hat \sigma _{lim}^{(1)} (k) \cos kx \, , \quad
\hat \sigma _{lim}^{(1)}(k)=2\pi \left [ \frac{e^{-|k|}}{1-e^{-|k|}}- \frac{e^{\frac{|k|}{2}}}{2\sinh \frac{|k|}{2}\cosh k} \right ] \, .
\ee
Going back to the coordinate space, we have
\ba
&& g \rightarrow +\infty \quad \Rightarrow \frac{d}{du} M(u,v) \rightarrow \frac{1}{2}[  \sigma _{lim}^{(1)} (u-v) +  \sigma _{lim}^{(1)} (u+v) ] \, , \nonumber \\
&& \sigma _{lim}^{(1)}(u)=- \frac{1}{4}\Bigl [ \psi \left (1-i\frac{u}{4} \right )+\psi \left (1+i\frac{u}{4} \right )- \psi \left (\frac{1}{2}-i\frac{u}{4} \right ) -
\psi \left (\frac{1}{2}+i\frac{u}{4} \right ) + \frac{2\pi}{\cosh \frac{\pi}{2}u} \Bigr ] \Rightarrow \nonumber \\
&& M(u,v)=-\frac{i}{2} \ln \frac{\Gamma \left (1-i\frac{u-v}{4} \right ) \Gamma \left (\frac{1}{2}+i\frac{u-v}{4} \right ) \Gamma \left (1-i\frac{u+v}{4} \right )\Gamma \left (\frac{1}{2}+i\frac{u+v}{4} \right )}{\Gamma \left (1+i\frac{u-v}{4} \right ) \Gamma \left (\frac{1}{2}-i\frac{u-v}{4} \right ) \Gamma \left (1+i\frac{u+v}{4} \right )\Gamma \left (\frac{1}{2}-i\frac{u+v}{4} \right )}- \nonumber \\
&-& \frac{1}{2}\textrm{gd} \left ( \frac{\pi (u-v)}{2} \right ) - \frac{1}{2}\textrm{gd} \left ( \frac{\pi (u+v)}{2} \right ) \, .
\ea
Therefore, for the scattering phase $\Theta (u,v)$ we have the expression \cite {BK}
\be
g\rightarrow +\infty \quad \Rightarrow \quad \Theta (u,v)\rightarrow -i \ln \frac{\Gamma \left (1-i\frac{u-v}{4} \right ) \Gamma \left (\frac{1}{2}+i\frac{u-v}{4} \right )}{\Gamma \left (1+i\frac{u-v}{4} \right ) \Gamma \left (\frac{1}{2}-i\frac{u-v}{4} \right )}-\textrm{gd} \left ( \frac{\pi (u-v)}{2} \right ) \, ,
\ee
which depends only on the difference of the rapidities.

\medskip

We now rescale the rapidity $u=\sqrt{2}g \bar u$ and then send $g\rightarrow +\infty$,
with $\bar u$ fixed. It is easier to compute the double derivative of the scattering factor $\Theta (u,v) $, since it  depends on the density $\sigma (u;v)$ only:
\be
\frac{d}{du}\frac{d}{dv} \Theta (u,v)= \frac{d}{dv}\sigma (u;v) - \frac{d}{du}\sigma (v;u) \label {theta-sigma} \, .
\ee
On the other hand, the function $\frac{d}{d\bar x}\sigma (u;x)$ is written (at the leading order $g^0$) as
\be
\frac{d}{d\bar x}\sigma (u;x)\cong \int _{0}^{+\infty} \frac{d\bar t}{\sqrt{2}g} \cos \bar t \bar u \left [ \frac{d}{d\bar x} \Gamma _-(\bar t;\bar x) - \frac{d}{d\bar x} \Gamma _+(\bar t;\bar x) - \frac{2e^{-\frac{\bar t}{\sqrt{2}g}}}{1-e^{-\frac{\bar t}{\sqrt{2}g}}} \bar t \sin \bar t \bar x  \right ] \label {sigma-gamma} \, ,
\ee
where the functions $\frac{d}{d\bar x}\Gamma (\bar t;\bar x)$ satisfy the integral equation, valid for $|\bar u|\leq 1$:
\be
\int _{0}^{+\infty} d \bar t \left [ e^{i\bar t \bar u} \frac{d}{d\bar x} \Gamma _- (\bar t;\bar x) - e^{-i\bar t \bar u} \frac{d}{d\bar x} \Gamma _+ (\bar t;\bar x)\right ] =  \int _{0}^{+\infty} d \bar t e^{i\bar t \bar u}  \frac{\bar t \sin \bar t \bar x}{\sinh \frac{\bar t}{2\sqrt{2}g}} \cong 2 \sqrt{2}g \frac{\bar x}{\bar x^2-\bar u^2} \, ,
\label {gamma-eq}
\ee
We set $\Gamma _+(\bar t; \bar x)= \int dk \cos k\bar t \, \tilde \Gamma (k;\bar x)$, $\Gamma _-(\bar t; \bar x)= -\int dk \sin k\bar t \, \tilde \Gamma (k;\bar x)$ and solve (\ref {gamma-eq}):
\ba
\frac{d}{d\bar x} \tilde \Gamma (k;\bar x)&=& - \sqrt{2}g [\delta (k-\bar x) - \delta (k+\bar x) ] + O(1/g) \, , \quad |\bar x| < 1 \label {solGamma1} \\
\frac{d}{d\bar x} \tilde \Gamma (k;\bar x)&=&- \frac{g}{\pi} \left [  \frac{ \left ( \frac{1+k}{1-k} \right )^{1/4} \left ( \frac{\bar x-1}{\bar x+1} \right )^{1/4}}{\bar x-k} + \frac{ \left ( \frac{1+k}{1-k} \right )^{1/4} \left ( \frac{\bar x+1}{\bar x-1} \right )^{1/4}}{\bar x+k} \right ] +O(1/g) \, , \, |k|<1, |\bar x| > 1 \label {solGamma2}
\ea
while for $|\bar x|,|k|>1$ $\tilde \Gamma (k; \bar x)$ is exponentially small.
Plugging (\ref {solGamma1}, \ref {solGamma2}) into (\ref {sigma-gamma}), we find
the following behaviour at the leading order $g^0$:
\ba
&& \frac{d}{d\bar x}\sigma (\sqrt{2}g\bar u;\sqrt{2}g\bar x)= -\frac{1}{2} H(\bar u^2-1) H(\bar x^2-1) \Bigl [ \frac {\left ( \frac{\bar x-1}{\bar x+1} \right )^{1/4} \left ( \frac{\bar u-1}{\bar u+1} \right )^{1/4} +
\left ( \frac{\bar x+1}{\bar x-1} \right )^{1/4} \left ( \frac{\bar u+1}{\bar u-1} \right )^{1/4} }{\bar u+\bar x}   + \nonumber \\
&+& \frac { \left ( \frac{\bar x-1}{\bar x+1} \right )^{1/4} \left ( \frac{\bar u+1}{\bar u-1} \right )^{1/4} +
\left ( \frac{\bar x+1}{\bar x-1} \right )^{1/4} \left ( \frac{\bar u-1}{\bar u+1} \right )^{1/4} } {\bar x-\bar u}   \Bigr ]
\label {sigma-sol} \, ,
\ea
with $H(x)$ the Heaviside function. Inserting (\ref {sigma-sol}) into (\ref {theta-sigma}), we obtain the leading order
\be
\frac{d}{d\bar v} \frac{d}{d\bar w} \Theta (\sqrt{2} g \bar v , \sqrt{2} g \bar w)= \sqrt{2}g H(\bar v^2-1) H(\bar w^2-1) \frac{\left ( \frac{\bar v+1}{\bar v-1}\right )^{1/4} \left ( \frac{\bar w-1}{\bar w+1}\right )^{1/4} +
\left ( \frac{\bar v-1}{\bar v+1}\right )^{1/4} \left ( \frac{\bar w+1}{\bar w-1}\right )^{1/4} }{\bar v-\bar w }  .
\label {der2theta}
\ee
Result (\ref {der2theta}) agrees with corresponding formula coming from the scattering phase (2.34) of \cite {DoreyZhao}.

\section{Gluonic excitations}
\setcounter{equation}{0}

The excitations of gauge fields on GKP string correspond
to insertions of partons of the type $D^{l-1}_\bot F_{+\bot}$ or $\bar{D}^{l-1}_\bot \bar{F}_{+\bot}$. As for the ABA, the field $D^{l-1}_\bot F_{+\bot}$ is represented as a stack of $l+1$ roots $u_3$, $l$ $u_2$ and $l-1$ $u_1$ roots \cite{Basso} of the Beisert-Staudacher equations \cite{Beisert-Staudacher} (similarly, $\bar{D}^{l-1}_\bot \bar{F}_{+\bot}$ comes out
from the replacement $(u_1,u_2,u_3)\rightarrow(u_7,u_6,u_5)$). Then, these large $s$ equations (an operator with some fields $D^{l-1}_\bot F_{+\bot}$ on a sea of covariant derivatives $D_+$ ($u_4$ roots)) can take on the form
\be\label{bethe-scal}
\bullet\quad\quad 1=e^{-i p_{k}L'}\prod_{j=1}^s\ \m{S}^{(44)}(u_{k},u_{j})\ \prod_{m}\prod_{i=1}^{N_{m}}\ \m{S}^{(4g)}_{m}(u_{k},u_i^m)\\
\ee
\be\label{bethe-glu}
\bullet\quad\quad 1=\prod_{m}\prod_{i=1}^{N_{m}}\ \m{S}^{(gg)}_{lm}(u_k^l,u_i^m)\ \prod_{j=1}^s\ \m{S}^{(g4)}_{l}(u_k^l,u_j) \, ,
\ee
where $\m{S}^{(44)}(u_k,u_j)$ describes the scattering involving two type-4 Bethe roots, while $\m{S}^{(4g)}_{m}(u_k,u_i^m)$ that of a type-4 root colliding with a gluonic stack (of length $m$ and real centre $u_i^m$)
and finally $\m{S}^{(gg)}_{lm}(u_k^l,u_i^m)$ represents the matrix for the scattering of gluonic bound states, in terms of real centres.
We are going to take into account a system composed of $Q$ gluonic bound states,
represented by stacks of length $m_k$ and real centre $\tilde{u}_k$, with $k=1,\dots ,Q$,
together with $L-2$ scalars, i.e. internal holes in the distribution of main roots:
the length $L'$ appearing in (\ref {bethe-scal}) equals $L+Q$.
As a matter of fact, in order to accommodate a gluonic excitation, a type-4 root has to be 'pulled away' from the sea, and a gluon substitute it, then. The state we are thus considering is characterised by $L+Q$ missing main roots in the sea, the vacuum corresponding to $L=2,Q=0$.

The counting function for a gluonic
stack, with real centre $u$ and length $l$, which collides only with scalars
and other gluonic excitations (centre rapidity $\ti{u}_k$, length $m_k$) is
\ba\label{Zg}
Z_g(u|l) = \sum_{k=1}^Q \tilde \chi(u,\ti{u}_k|l,m_k)- \int_{-\infty}^{+\infty}\frac{dv}{2\pi} \chi(v,u|l)\,\frac{d}{dv} [Z^{(Qg)}(v)-2 L^{(Qg)}(v) ]
-\sum_{h=1}^L \chi(x_h,u|l) \, ,
\ea
whereas the counting function (\ref{Z}) adapted to the case at hand, including the $Q$ gluonic stacks, takes on the form:
\ba
Z^{(Qg)}(v)&=& (L+Q) \Phi (v) + \int_{-\infty}^{+\infty}\frac{dw}{2\pi} \phi (v,w) \frac{d}{dw} [ Z^{(Qg)}(w)-2 L^{(Qg)}(w) ] + \nonumber\\
&+& \sum _{h=1}^L \phi (v,x_h) + \sum_{k=1}^Q\chi(v,\tilde u_k|m_k)  \, . \label {Z1g}
\ea
In addition to definitions (\ref{Phi}, \ref {phi}), we introduced
\ba
&&\chi(v,u|l)\equiv \chi _0(v-u|l+1)+\chi_H(v,u-\frac{il}{2})+\chi_H(v,u+\frac{il}{2}) \, , \\
&&\ti{\chi}(u,v|l,m)\equiv \chi_0(u-v|l+m)-\chi_0(u-v|l-m)+2\sum_{\gamma=1}^{l-1}\chi_0(u-v|l+m-2\gamma)\, ,
\ea
where we split the function $\chi$ into its one-loop and higher than one loop parts, respectively
\ba
\bullet \chi _0(u|l)\equiv 2\ \arctan\frac{2u}{l}=i\ln \frac{il +2u}{il -2u} \qquad\qquad
\bullet\chi_H(u,v)\equiv i\ln\left(\frac{1-\frac{g^2}{2x^-(u)x(v)}}{1-\frac{g^2}{2x^+(u)x(v)}}\right)\ \ .
\ea
Equation (\ref {Z1g}) suggests that $Z^{(Qg)}(v)$ satisfies the nonlinear integral equation
\be
Z^{(Qg)}(v)=F^{(Qg)}(v)+2 \int _{-\infty}^{+\infty} dw \, G(v,w) L ^{(Qg)}(w) \, ,
\ee
where the function $F^{(Qg)}(v)$ is written as
\be
F^{(Qg)}(v)=(L+Q) \tilde P(v) + \sum _{h=1}^L R(v,x_h) + \sum_{k=1}^Q T(v,\tilde u_k |m_k) \, ,
\ee
with $R(v,u)$ and $\tilde P(v)$ solutions of (\ref {tildePeq}) respectively, and $T(v,\tilde u|m)$ equals
\be
T(v, \tilde u |m) = \chi(v,\tilde u |m) - \int_{-\infty}^{+\infty}dw  \, G(v,w) \chi (w, \tilde u |m) \, .  \label {T}
\ee
We are interested in the scattering factors involving gluons, which correspond to $l=1$ stacks. Therefore, we restrict to this case and, for clarity's sake,
denote the scalar-scalar factor as $S^{(ss)}(x_h,x_{h'})=-\textrm{exp} [-i\Theta (x_h,x_{h'})]$. The quantisation conditions for holes and gluons are
\ba
&&\bullet   \ (-1)^{L-1}=e^{-i Z^{(Qg)}(x_h)}= e^{iRP(x_h)} \prod _{h'=2, h'\not=h}^{L-1} \left (- S^{(ss)}(x_h,x_{h'}) \right ) \prod _{j=1}^Q S^{(sg)}(x_h, \tilde u_j) \label {quacon1} \\
&&\bullet (-1)^{Q-1}=e^{-iZ_g(\tilde u_k|1)}=   e^{iRP^{g}(\tilde u_k)} \prod _{j=1, j\not=k}^{Q} \left (- S^{(gg)}(\tilde u_k,\tilde u_j) \right ) \prod _{h'=2}^{L-1}S^{(gs)}(\tilde u_k, x_{h'}) \label {quacon2} \, .
\ea
In order to gain $S^{(gg)}$ we consider
(\ref{Zg}) when the system is composed just of two gluons ($Q=2$ and $l=1$) with rapidities $\ti u_1$ and $\ti u_2$ and no internal holes are present ($L=2$):
\ba
Z_g(u|1)= -4\int_{-\infty}^\infty \frac{dv}{2\pi}\, \chi(v,u|1) \, \frac{d}{dv} \ti{P}(v)+
\int_{-\infty}^\infty \frac{dv}{\pi}\, \frac{d L^{(2g)}}{dv}(v) \, T(v,u|1)- \nonumber\\
-\sum_{h=1}^2 T(x_h,u|1)
+\sum_{k=1}^2 \left[\ti{\chi}(u,\ti{u}_k|1,1) - \int_{-\infty}^\infty \frac{dv}{2\pi}\, \chi(v,u|1) \, \frac{d}{dv} T(v,\ti{u}_k|1) \right] \ . \label {Zg2}
\ea
Via quantisation condition (\ref {quacon2}), from (\ref {Zg2}) we can identify the momentum $P^{(g)}(u)$ of a gluon with rapidity $u$
\ba\label{momentumGluon}
T(x_1,u|1)+T(x_L,u|1)
-\int_{-\infty}^\infty \frac{dv}{\pi}\, \frac{d L^{(Qg)}}{dv}(v) \, T(v,u|1) - \sum _{h=2}^{L-1} \tilde P(x_h)  + \\
+  2 \int _{-\infty}^{+\infty} \frac{dv}{2\pi} \chi (v,u|1) \frac{d}{dv}\tilde P(v) +\sum _{k=1}^Q \int _{-\infty}^{+\infty} \frac{dv}{2\pi} \chi (v,\tilde u_k|1) \frac{d}{dv}\tilde P(v) \simeq R \cdot P^{(g)}(u) \nonumber
\ ,
\ea
with effective length $R\simeq 2\ln s$ and the scattering phase between gluons with rapidities $u$ and $\ti u$:
\ba\label{S_gg}
&& i\ln \left (-S^{(gg)}(u,\tilde u)\right )=\ti{\chi}(u,\ti{u}|1,1) - \int_{-\infty}^\infty \frac{dv}{2\pi}\, \chi(v,u|1) \, \frac{d}{dv} T(v,\ti{u}|1) - \int_{-\infty}^\infty \frac{dv}{2\pi} \chi (v, u|1)  \frac{d}{dv} \tilde P(v) + \nonumber \\
&+& \int_{-\infty}^\infty \frac{dv}{2\pi} \chi (v, \tilde u|1)  \frac{d}{dv} \tilde P(v) = \ti{\chi}(u,\ti{u}|1,1) - \int_{-\infty}^{+\infty} \frac{dv}{2\pi}\, [ \chi(v,u|1)
+\Phi (v)]\, \frac{d}{dv}[ \chi(v,\ti{u}|1)+\Phi (v) ] +  \nonumber \\
&+& \int_{-\infty}^{+\infty} \frac{dv}{2\pi}\int_{-\infty}^{+\infty} \frac{dw}{2\pi}\, \,  [ \chi(v,u|1)
+\Phi (v)] \left [ \frac{d}{dv} \frac{d}{dw} \Theta (v,w) \right ] \,  [ \chi(w,\tilde u|1)
+\Phi (w)]  \, ,  \label {Sgg}
\ea
where $\Theta$ (\ref {ThetaMrel}) enters the hole-hole scattering phase.
This expression at one loop reduces to
\be
S^{(gg)}(u,\ti{u})=-
\frac{\Gamma\left(1+i(u-\ti{u})\right)}{\Gamma\left(1-i(u-\ti{u})\right)}
\frac{\Gamma\left(\frac{3}{2}-iu\right)}{\Gamma\left(\frac{3}{2}+iu\right)}
\frac{\Gamma\left(\frac{3}{2}+i\ti{u}\right)}{\Gamma\left(\frac{3}{2}-i\ti{u}\right)}\ \ , \label {Sgg2}
\ee
which agrees with relations (7) and (11) of \cite {BSV}. Moreover, the gluonic counting function (\ref{Zg}) allows us to retrieve the scattering matrix between a gluon ($l=1$ and rapidity $\ti{u}$) and a scalar excitation
(internal hole with rapidity $x_h$), once properly rewritten after fixing $Q=1$ and $L=3$:
\ba
Z_g(u|1)= -4\int_{-\infty}^\infty \frac{dv}{2\pi}\, \chi(v,u|1) \, \frac{d}{dv} \ti{P}(v)+
\int_{-\infty}^\infty \frac{dv}{\pi}\, \frac{d L^{(1g)}}{dv}(v) \, T(v,u|1)+ \nonumber\\
-\sum_{h=1}^3 T(x_h,u|1)
+\left[\ti{\chi}(u,\ti{u}|1,1) - \int_{-\infty}^\infty \frac{dv}{2\pi}\, \chi(v,u|1) \, \frac{d}{dv} T(v,\ti{u}|1) \right] \, .
\ea
Therefore, the gluon-scalar scattering phase reads
\ba
&& i\ln \left (S^{(gs)}(\ti{u},x_h)\right ) = -  \chi (x_h,\ti{u}|1) - \Phi (x_h) + \int _{-\infty}^{+\infty} \frac{dw}{2\pi} \left [ \frac{d}{dw}\Theta (x_h, w) \right ] \left ( \chi (w,\ti{u}|1) +\Phi (w) \right ) = \nonumber  \\
&& = \int _{-\infty}^{+\infty} \frac{dw}{2\pi} \left [ \frac{d}{dw}\Theta (x_h, w)-2\pi \delta (x_h-w) \right ] \left ( \chi (w,\ti{u}|1) +\Phi (w) \right )
=-i\ln \left ( S^{(sg)}(x_h,\tilde u) \right )
\label {S-gs} \, ;
\ea
at one loop, it becomes
\be
S^{(gs)}(\ti u,x_h)=
\frac{\Gamma\left(1+i(\ti u-x_h)\right)}{\Gamma\left(1-i(\ti u-x_h)\right)}
\frac{\Gamma\left(\frac{1}{2}+ix_h\right)}{\Gamma\left(\frac{1}{2}-ix_h\right)}
\frac{\Gamma\left(\frac{3}{2}-i\ti{u}\right)}{\Gamma\left(\frac{3}{2}+i\ti{u}\right)}
= [S^{(sg)}(x_h,\tilde u)]^{-1} \, .
\ee
Finally, we consider scattering between $F_{+\bot}$, with rapidity $\ti u_1$, and $\bar{F}_{+\bot}$, with rapidity $\ti{\bar u}_1$.
We stick to $L=2$ and consider $Z_g$, counting function of $F_{+\bot}$ and $Z^{(g\bar g)}$, counting function of scalars in the presence of
$F_{+\bot}$ and $\bar{F}_{+\bot}$:
\ba
Z_g(u|1) &=& \tilde \chi(u,\ti{u}_1|1,1)- \int_{-\infty}^{+\infty}\frac{dv}{2\pi} \chi(v,u|1)\,\frac{d}{dv} [Z^{(g\bar g)}(v)-2 L^{(g\bar g)}(v) ]
-\sum_{h=1}^2 \chi(x_h,u|1) \, , \label {Zgu1}\\
Z^{(g\bar g)}(v)&=&4 \tilde P(v) + \sum _{h=1}^2 R(v,x_h) + T(v,\tilde u_1 |1)+  T(v,\ti {\bar u}_1 |1) + NL(v) \ \label {ZQg} \, .
\ea
Plugging (\ref {ZQg}) into (\ref {Zgu1}), computing $Z_g$ in $u=\tilde u_1$ and neglecting non linear terms, we obtain
\be
Z_g (\tilde u_1 | 1) = - \int_{-\infty}^{+\infty}\frac{dv}{2\pi}  \chi(v,\tilde u_1|1) \left [ 4\frac{d}{dv} \tilde P(v) +
\frac{d}{dv} T(v, \ti {\bar u}_1 |1) \right ]
- T(x_1, \tilde u_1|1) - T(x_L, \tilde u_1|1) \, .
\ee
The quantisation condition reads $1=e^{-i Z_g (\tilde u_1 | 1) }=e^{iRP^{(g)}(\tilde u_1)} S^{(g\bar g)}(\tilde u_1, \ti {\bar u}_1)$,
from which, using (\ref {momentumGluon}, \ref {S_gg}) we obtain
\be
i\ln S^{(g\bar g)} (\tilde u_1, \ti {\bar u}_1) = i\ln \left ( - S^{(gg)} (\tilde u_1, \ti {\bar u}_1) \right ) -
\chi (\tilde u_1, \ti {\bar u}_1 |1,1)
\Rightarrow
S^{(g\bar g)} (\tilde u_1, \ti {\bar u}_1)= S^{(gg)} (\tilde u_1, \ti {\bar u}_1)\frac{\tilde u_1-\ti {\bar u}_1-i}{\tilde u_1-\ti {\bar u}_1+i} \, .
\ee
Analogously, considering the counting function of $\bar{F}_{+\bot}$ together with $Z^{(g\bar g)}$, we find $S^{(\bar g g)}( \ti {\bar u}_1, \tilde u_1)=
[S^{(g\bar g)}(\tilde u_1, \ti {\bar u}_1)]^{-1}$. Eventually, we also obtain $S^{(\bar gs)}(u,v)=S^{(gs)}(u,v)$.

\medskip
\noindent
{\bf Perturbative strong coupling gluon-gluon scattering}

We want to study the gluonic scattering matrix (\ref {Sgg}) in the limit $g\rightarrow +\infty$, with
$u=\bar u \sqrt{2}g$, $\tilde u=\bar {\tilde u } \sqrt{2}g$, $\bar u$, $\bar {\tilde u } $ fixed and
$ \bar u ^2 <1$,  $\bar {\tilde u } ^2 <1$.
We have
\be
i\ln \left (-S^{(gg)}(u,\tilde u)\right) = \mathcal{I}_1+\mathcal{I}_2+\mathcal{I}_3 \, ,
\ee
where
\be
 \mathcal{I}_1=\tilde \chi (u,\tilde u|1,1)=
 -2 \arctan \sqrt{2}g (\bar {\tilde u }-\bar u) =
 -\pi \textrm{sgn}(\bar {\tilde u }-\bar u) + \frac{\sqrt{2}}{g(\bar {\tilde u }-\bar u)}+ O(1/g^3) \, ,  \label {i1}
\ee
\be\label{i2}
\mathcal{I}_2= -2 \arctan \left [ \frac{g(\bar {\tilde u }-\bar u)}{\sqrt{2}} \right ] - 2 \arctan [g(\bar {\tilde u }-\bar u)\sqrt{2}] +4 \arctan  \left [ \frac{2 \sqrt{2}g(\bar {\tilde u }-\bar u)}{3} \right ]
= O(1/g^3) \, .
\ee
For what concerns the last term $\mathcal{I}_3$ in the right hand side of (\ref {Sgg}), we find convenient to perform the change of variables
$v=\sqrt{2} g \bar v$, $w=\sqrt{2} g \bar w$
\ba
\mathcal{I}_3 \cong \int_{-\infty}^{+\infty} \frac{d\bar v}{2\pi} \int_{-\infty}^{+\infty} \frac{d\bar w}{2\pi}
\frac{1}{\sqrt{2}g \bar v-\sqrt{2}g\bar u } \frac{1}{\sqrt{2}g \bar w-\sqrt{2}g \bar {\tilde u }}
\frac{d}{d\bar v} \frac{d}{d\bar w} \Theta (\sqrt{2} g \bar v , \sqrt{2} g \bar w) \, .  \label {i3}
\ea
Plugging formula (\ref {der2theta}) into (\ref {i3}) and performing the integrations we arrive at
\be
\mathcal{I}_3= \frac{1}{2\sqrt{2}g (\bar u - \bar {\tilde u } )} \left [ 2 - \left ( \frac{1+\bar u}{1- \bar u}\right )^{\frac{1}{4}} \left ( \frac{1-\bar {\tilde u }}{1+\bar {\tilde u } }\right )^{\frac{1}{4}}  - \left ( \frac{1-\bar u}{1+ \bar u}\right )^{\frac{1}{4}} \left ( \frac{1+\bar {\tilde u }}{1-\bar {\tilde u } }\right )^{\frac{1}{4}} \right ] + O(1/g^2) \label {i3fin} \, .
\ee
Now, summing up (\ref {i1}, \ref {i2}, \ref {i3fin}) we obtain the final result for the gluon-gluon scattering matrix at the order $O(1/g)$
\be
S^{(gg)}(u,\tilde u)= \textrm{exp} \left [ \frac{i}{\sqrt{2}g(\bar u - \bar {\tilde u })}\left ( 1+
\frac{1}{2}\left ( \frac{1+\bar u}{1- \bar u}\right )^{\frac{1}{4}} \left ( \frac{1-\bar {\tilde u }}{1+\bar {\tilde u } }\right )^{\frac{1}{4}}  + \frac{1}{2}\left ( \frac{1-\bar u}{1+ \bar u}\right )^{\frac{1}{4}} \left ( \frac{1+\bar {\tilde u }}{1-\bar {\tilde u } }\right )^{\frac{1}{4}}  +O(1/g^2) \right )\right ] \, , \label {S11fin}
\ee
which agrees with the result coming from (7), (15) and (16) of \cite {BSV}.

\section{Fermionic Excitations}
\setcounter{equation}{0}

To parametrise the dynamics of a fermionic excitation, the rapidity to look at is
actually $x$, which is related to the Bethe rapidity $u$ via the Zhukovski map $u(x)=x+\frac{g^2}{2x}$. To properly invert the Zhukovski map for the complete range of values of $x$, we need to glue two $u$-planes together, each corresponding to a Riemann sheet. The two sheets are related to two distinct regimes of the fermionic excitations \cite{Basso}: large fermions, embedded in Beisert-Staudacher equations \cite{Beisert-Staudacher} as
$u_3$ roots, which do carry energy and momentum even at one-loop; small fermions, corresponding
to $u_1$ roots, which couple to main root equations just at higher loops. The function $x(u)$ can be analytically continued from the $u_3$ Riemann sheet to the $u_1$-sheet by means of the map
$x(u_3) \longrightarrow (g^2)/(2x(u_1))$, and Beisert-Staudacher equations are invariant under this exchange $u_3 \leftrightarrow u_1$, provided we modify the spin-chain length (see \cite{Beisert-Staudacher} for details). Exactly the same reasoning applies for anti-fermions by replacing $u_3\rightarrow u_5$ and $u_1\rightarrow u_7$: turning on $u_3$ ($u_1$) roots means exciting fermionic fields $\Psi_+$, while
$u_5$ ($u_7$) corresponds to $\bar{\Psi}_+$. Hence, we can extend (\ref{bethe-scal},\,\ref{bethe-glu}) to include $N_F$ large fermions $u^F=u_3$, of physical rapidities $x_{j}^F=x(u_{j}^F)$ with the arithmetic square root for $x(u)=(u/2)\left[1+\sqrt{1-(2g^2)/u^2}\right]$, and $n_f$ small fermions $u^f=u_1$, of rapidities $x_{j}^f= (g^2)/(2 x(u_{j}^f))$:
\ba
\bullet\ &1&=e^{-i p_k L'}\prod_{j\neq k}^s\m{S}^{(44)}(u_k,u_j)\prod_{j=1}^{N_F}\m{S}^{(4F)}(u_k,u_{j}^F)\prod_{j=1}^{n_f}\m{S}^{(4f)}(u_k,u_{j}^f)
  \prod_m \prod_{j=1}^{N_m} \m{S}^{(4g)}_m(u_k, u_j^m)\\
\bullet\ &1&=\prod_{j=1}^s\m{S}^{(F4)}(u_{k}^F,u_j) \prod_m \prod_{j=1}^{N_m} \m{S}^{(Fg)}_m(u_k^F, u_j^m)\\
\bullet\ &1&=\prod_{j=1}^s\m{S}^{(f4)}(u_{k}^f,u_j) \prod_m \prod_{j=1}^{N_m} \m{S}^{(fg)}_m(u_k^f, u_j^m)\\
\bullet\ &1&=\prod_{m}\prod_{i=1}^{N_{m}}\ \m{S}^{(gg)}_{lm}(u_k^l,u_i^m)\ \prod_{j=1}^s\ \m{S}^{(g4)}_{l}(u_k^l,u_j)
  \prod_{j=1}^{N_F}\m{S}^{(gF)}_{l}(u_k^l,u_{j}^F)\prod_{j=1}^{n_f}\m{S}^{(gf)}_{l}(u_k^l,u_{j}^f) \,\,\, ,
\ea
where $L'=L+N_F+Q$ and, in addition to previously defined matrices, we introduce the scattering matrices describing the collision between a type-4 root and a large fermion $\m{S}^{(4F)}(u_k,u_{j}^F)$ or small fermion $\m{S}^{(4f)}(u_k,u_{j}^f)$
together with their inverses (respectively $\m{S}^{(F4)}(u_{k}^F,u_j)$, $\m{S}^{(f4)}(u_{k}^f,u_j)$);
the matrices $\m{S}^{(gF)}_{l}(u_k^l,u_{j}^F)$, $\m{S}^{(Fg)}_{m}(u_k^F,u_{j}^m)$ (or $\m{S}^{(gf)}_{l}(u_k^l,u_{j}^f)$ and $\m{S}^{(fg)}_{m}(u_k^f,u_{j}^m)$) describe the scattering involving gluonic stacks and large (small) fermions. Explicitly, these scattering matrices over the GKP vacuum are listed here (the definition $\chi_F (v,u)=\chi_0(v-u|1)+\chi_H(v,u)$ is used):\\
$\bullet$ large (anti)fermion-large (anti)fermion: $S^{(FF)}(u,u')=S^{(F\bar F)}(u,u')=S^{(\bar FF)}(u,u')=S^{(\bar F\bar F)}(u,u')$,
\be
i\log S^{(FF)}(u,u')=-\int \frac{dv}{2\pi}\frac{dw}{2\pi}\left[\chi_F (v,u)+\Phi(v)\right]\frac{d}{dv}\left(2\pi\delta(v-w)-\frac{d\Theta}{dw}(v,w)\right)
\left[\chi_F (w,u')+\Phi(w)\right]
\ee
$\bullet$ large (anti)fermion-small (anti)fermion: $S^{(Ff)}(u,u')=S^{(F\bar f)}(u,u')=S^{(\bar Ff)}(u,u')=S^{(\bar F\bar f)}(u,u')$,
\be
i\log S^{(Ff)}(u,u') =\int \frac{dv}{2\pi}\frac{dw}{2\pi}\left[\chi_F (v,u)+\Phi(v)\right]\frac{d}{dv}\left(2\pi\delta (v-w)-\frac{d\Theta}{dw}(v,w)\right)
\chi_H(w,u')
\ee
$\bullet$ scalar-large (anti)fermion: $S^{(sF)}(u,u')=S^{(s\bar F)}(u,u')$,
\ba
i \log S^{(sF)}(u,u')=
-\int \frac{dv}{2\pi}\,\left[\frac{d\Theta}{dv}(u,v)-2\pi\delta(u-v)\right]\,\left(\chi_F (v,u')+\Phi(v)\right)
\ea
$\bullet$ gluonic stack-large (anti)fermion: $S^{(gF)}_{l}(u,u')=S^{(\bar g\bar F)}_{l}(u,u')$, $S^{(g\bar F)}_{l}(u,u')=S^{(\bar gF)}_{l}(u,u')$,
\ba
&&i\log(- S^{(gF)}_{l}(u,u')) = \chi_0(u-u'|l)+ i\log S^{(\bar gF)}_{l}(u,u')=\\
&&=\chi_0(u-u'|l)
-\int \frac{dv}{2\pi}\frac{dw}{2\pi}\left[\chi(v,u|l)+\Phi(v)\right]\frac{d}{dv}\left(2\pi\delta (v-w)-\frac{d\Theta}{dw}(v,w)\right)
\left[\chi_F (w,u')+\Phi(w)\right] . \nonumber
\ea
We checked that unitarity holds, i.e. that $S^{(sF)}(u,u')=[S^{(Fs)}(u',u)]^{-1}$, $S^{(gF)}_{l}(u,u')=[S^{(Fg)}_{l}(u',u)]^{-1}$ and so on.
By virtue of the map between small and large (anti)fermions, from the expressions above
we recover the corresponding ones for small (anti)fermions by replacing
$\chi_F (v,u)+\Phi (v) \, \longrightarrow\, -\chi_H (v,u)$.
Eventually, we note that all these scattering phases depend only on the 'basic' scalar-scalar one, except known functions.

\noindent
{\bf Note Added}: When completing to write this work, \cite{Basso:2013pxa} on scalars appeared (one day in advance). In fact, \cite{Basso:2013pxa} focuses on scalars giving for them the complete S-matrix (namely the $g$ dependent scalar-scalar (pre)factor, which is also presented in this letter, times a matrix fixed by the $O(6)$ symmetry, as in the $O(6)$ NLSM). In this letter we derive all the (pre)factors concerning the other fields, {\it i.e.} gluons and fermions.

\medskip

{\bf Acknowledgements}
We enjoyed discussions with B. Basso, D. Bombardelli, N.Dorey and P. Zhao. This project was  partially supported by INFN grants IS  FI11 and PI14, the Italian
MIUR-PRIN contract 2009KHZKRX-007, the ESF Network 09-RNP-092 (PESC) and the MPNS--COST Action MP1210.


\end{document}